# Demonstration of Dissipative Quasihelical Edge Transport in Quantum Anomalous Hall Insulators

Shu-Wei Wang[1,*], Di Xiao[2], Ziwei Dou[1], Moda Cao[1], Yi-Fan Zhao[2], Nitin Samarth[2], Cui-Zu Chang[2], Malcolm R. Connolly[3], and Charles G. Smith[1]

[1]*Cavendish Laboratory, Department of Physics, University of Cambridge, Cambridge, CB3 0HE, UK*

[2]*Department of Physics, The Pennsylvania State University, University Park, Pennsylvania 16802, USA*

[3]*Department of Physics, Imperial College London, London, SW7 2AZ, UK*

[*]Corresponding author: sww38@cam.ac.uk

Doping a topological insulator (TI) film with transition metal ions can break its time-reversal symmetry and lead to the realization of the quantum anomalous Hall (QAH) effect. Prior studies have shown that the longitudinal resistance of the QAH samples usually does not vanish when the Hall resistance shows a good quantization. This has been interpreted as a result of the presence of possible dissipative conducting channels in magnetic TI samples. By studying the temperature- and magnetic field-dependence of the magnetoresistance of a magnetic TI sandwich heterostructure device, we demonstrate that the predominant dissipation mechanism in thick QAH insulators can switch between non-chiral edge states and residual bulk states in different magnetic field regimes. The interactions between bulk states, chiral edge states, and non-chiral edge states are also investigated. Our study provides a way to distinguish between the dissipation arising from the residual bulk states and non-chiral edge states, which is crucial for achieving true dissipationless transport in QAH insulators and for providing deeper insights into QAH-related phenomena.

The quantum Hall (QH) effect discovered by von Klitzing *et al.* in 1980 marked the first observation of the topological phase in solid state systems [1,2]. In the QH effect in a two-dimensional electron system (2DES), electrons move along chiral edge channels and give rise to a vanishing longitudinal resistance ($\rho_{xx}$) and a quantized Hall resistance ($\rho_{xy}$) at $h/\nu e^2$, where $h$ is the Planck constant, $e$ is the elementary charge, and $\nu$ is an integer known as the filling factor that represents the number of chiral edge channels [3]. However, the QH effect typically requires a strong external magnetic field, which hampers the development of technologies based on the ballistic transport of the QH edge states.

The possibility of a QH effect without the need of external magnetic fields was envisioned by Haldane in 1988 [4], and the Haldane model was later adopted to propose the existence of the quantum spin Hall (QSH) effect [5–7]. Following the theoretical prediction [7], the QSH effect was realized in the HgTe/CdTe quantum well [8], which has a topologically non-trivial band structure due to the intrinsic strong spin orbit coupling (SOC) of the material [7]. In the QSH effect, the four-terminal longitudinal resistance is quantized to $h/2e^2$ at zero magnetic field [7,8], and the edge channel in the QSH effect is known as a helical edge state, which is protected by time-reversal symmetry (TRS) and is thus immune to backscattering from non-magnetic perturbations.

The model for the QSH effect was later generalized to three-dimensional (3D) systems [9–12], and the materials with topologically non-trivial band structures were named "topological insulators" (TIs) [10]. Theoretical studies predicted that breaking TRS of TIs with magnetic perturbations can give rise to the quantum anomalous Hall (QAH) effect, in which only a spin-polarized chiral edge channel propagates along the edge of the sample without the need of external magnetic fields [13,14]. By doping transition metal ions into TI films, the QAH effect has been successfully observed in Cr- and V-doped $(Bi,Sb)_2Te_3$ samples [15,16]. Nevertheless, it was found that the QAH effect observed in experiments usually shows a non-vanishing $\rho_{xx}$ while $\rho_{xy}$ is quantized at $h/e^2$ at zero magnetic field, suggesting the existence of dissipative conducting channels [15–21]. The possible reasons for the dissipation include variable range hopping [15,22], non-chiral edge states [17,20,23], and thermally activated bulk and surface carriers [19,21]. It has been pointed out that the side surface of a thicker magnetic TI sample may not be gapped and is likely to accommodate quasihelical edge states, which originate from the topological surface states of the TI but are no longer protected by TRS owing to the in-plane magnetization on the side surface [17,20,23]. The existence of quasihelical edge states is also speculated to be the reason that ultralow temperatures are needed for the observation of the QAH effect [20]. Although the quasihelical edge states were used to explain the non-zero $\rho_{xx}$ and the low critical temperature of the QAH effect, the behavior of these dissipative edge states remains poorly understood. Recently, the QAH state with higher critical temperatures was realized in modulation-doped magnetic TI heterostructures [24–27]. These magnetic TI heterostructures not only provide a better platform for the study of the QAH effect but also enable the exploration of novel physics (e.g., axion physics) that is not available in the uniformly doped magnetic TI films [25,26,28]. In this Letter, we systematically study the behavior of the quasihelical edge states and the coexistence of chiral edge states, non-chiral edge states, and bulk states in a QAH system by investigating the transport properties of a magnetic TI sandwich heterostructure. We show that the predominant dissipation mechanism in a thick QAH insulator can switch between non-chiral edge states and residual bulk states in different magnetic field regimes.

The sample used in our study was grown on a 0.5 mm thick heat-treated SrTiO$_3$ (111) substrate in a molecular beam epitaxy (MBE) chamber. The sample consists of a five quintuple-layer (QL) of (Bi,Sb)$_2$Te$_3$ region sandwiched by two 3 QL Cr-doped (Bi,Sb)$_2$Te$_3$ regions, as shown in Fig. 1(a). A layer of silver paste was used to apply a back-gate voltage ($V_g$) onto the SrTiO$_3$ substrate to tune the Fermi level of the magnetic TI. To avoid contamination from lithographic processing, the sample was scratched into a Hall bar (see more information in the Supplemental Material [29]). The transport measurements were carried out in a Bluefors dilution refrigerator using a 1 nA excitation current with perpendicular magnetic fields up to 12 T.

We first demonstrate the QAH effect in this magnetic TI sandwich sample. A prerequisite for the realization of the QAH effect is that the Fermi level must be tuned into the exchange gap of the magnetic TI sample, which can be achieved by adjusting $V_g$. The $V_g$ dependence of the minimum $\rho_{xx}$ and maximum $\rho_{xy}$ at the sample temperature $T_s = 16$ mK is shown in Fig. 1 (b). We note that the minimum $\rho_{xx}$ and maximum $\rho_{xy}$ appear at $\sim 90$ mT before reaching zero magnetic field in the demagnetization process. This is likely due to the demagnetization cooling effect that lowers the electron temperature and hence improves the QAH effect [19] (see more discussions in the Supplemental Material [29]). Although the lowest $\rho_{xx} \sim 0.0667$ $h/e^2$ and the highest $\rho_{xy} \sim$ 0.998 $h/e^2$ appear at $V_g$ = +200 V, the trend shows that $\rho_{xx}$ is still descending at $V_g$ = +200 V, suggesting that the charge neutrality point may be somewhere beyond $V_g$ = +200 V. Because of the charging effect of the SrTiO₃ substrate at high electric field, $V_g$ is not ramped beyond +200 V. Nevertheless, the nearly quantized $\rho_{xy}$ and the large Hall angle ~ 86.2º indicate that the Fermi level is located in the exchange gap and very close to the charge neutrality point (i.e., the minimum of the $\rho_{xx}$-$V_g$ curve). The small difference between the Fermi level and the charge neutrality point should not cause qualitative difference in our study (see more discussions in the Supplemental Material [29]). Thus, all the measurements hereafter are chosen to be performed at $V_g = V_g^{0}$ = +200 V. The magnetic field (μ₀H) dependence of $\rho_{xx}$ and $\rho_{xy}$ measured at $V_g^{0}$ is shown in Fig. 1 (c), in which $\rho_{xx}$ displays a butterfly shape while $\rho_{xy}$ shows a square hysteresis loop with the resistance quantized at $\pm h/e^2$.

To further investigate the chiral edge channels in the QAH effect, we perform three terminal measurements at $V_g = V_g^{0}$ and $T_s$ = 16 mK, as shown in Fig. 2(a). The current is passed from electrode 4 to electrode 6, and the magnetic field is swept between μ₀H = −2 T and +2 T. The four different configurations all exhibit a hysteresis loop in magnetic field sweeps, which is a manifestation of the chirality of the QAH edge states, as illustrated in Fig. 2(b). When the magnetization $M$ is negative (positive), the chiral edge states will travel along the direction of the green (orange) arrows. When the conduction is via the green path (i.e., $M$ < 0), ideally, no dissipation should occur before the chiral edge states reach electrode 6. Thus, electrodes 3, 2, and 1 should be at the same potential as electrode 4, while electrodes 5 and 6 should be grounded. In this case, $\rho_{46,41}$, $\rho_{46,42}$, and $\rho_{46,43}$ should be close to zero, and $\rho_{46,45}$ should be around $h/e^2$. On the other hand, when the conduction is via the orange path (i.e., $M$ > 0), electrode 5 should be at the same potential as electrode 4, while electrodes 1, 2, and 3 should be at the same potential as electrode 6 (i.e., being grounded), leading to $\rho_{46,41}$, $\rho_{46,42}$, and $\rho_{46,43}$ ~ $h/e^2$ while $\rho_{46,45}$ ~ 0. This behavior can be clearly observed in all four hysteresis loops in Fig. 2(a), in which the traces in the $M$ > 0 regime always differ from the traces in the $M$ < 0 regime by ~ $h/e^2$. These three-terminal measurements clearly demonstrate the chirality of the QAH edge states. However, the three-terminal resistance in Fig. 2(a) is only close to zero near μ₀H = 0 T and does not completely vanish throughout the entire well magnetized regime. Moreover, a slope in the magnetoresistance showing a deviation from $\rho_{xx}$ = 0 and $\rho_{xy}$ = $h/e^2$ can be observed in both Figs. 2(a) and 1(c), suggesting the existence of dissipative channels other than the chiral edge states even in the QAH regime. We note that the contact resistance of electrode 4 is not subtracted in the three-terminal measurements because the contact resistance of millimeter-size indium contacts is negligible [20,29].

To investigate the origin of these dissipative channels in the QAH regime, we perform non-local measurements to distinguish between the contributions from the bulk and the edge. The classical bulk contribution to the non-local signal can be estimated by the van der Pauw equation $\rho_{\mathrm{NL}}^{\mathrm{classical}}/\rho_{xx} \approx \exp(-\pi L/W)$, where $L$ is the distance between voltage probes and $W$ is the channel width [23]. For our device, $L/W \sim 6.5$ and hence $\exp(-\pi L/W) \approx 1.3{\times}10^{-9}$. However, the ratio between the measured non-local resistance and local resistance is found to be $\rho_{16,43}/\rho_{xx} \approx 3{\times}10^{-2}$, which is too high to be merely ascribed to the diffusive bulk contribution (see more discussions in the Supplemental Material [29]). This implies that the dissipation is not solely caused by the residual bulk states and may also be induced by some kind of non-chiral edge states [17,20,23]. Thus, we examine the $\mu_0 H$ dependence of non-local signals using two different configurations: $\rho_{16,43}$ and $\rho_{26,35}$. As shown in Fig. 2(c), both $\rho_{16,43}$ and $\rho_{26,35}$ still show hysteresis in the magnetic field sweeps, which further confirms the existence of dissipative edge states [17,20]. A possible candidate for the dissipative edge state is the aforementioned quasihelical edge state on the side surface of thick magnetic TI samples [23]. Since the sample used here is a sandwich heterostructure with a total thickness of 11 QL, the existence of such quasihelical edge states is expected [17,23]. The slope of the magnetoresistance in the fully magnetized regime in Figs. 1(c), 2(a), and 2(c) may also arise from the magnetic-field dependent behavior of these non-chiral edge states.

To scrutinize this observation, we perform high magnetic field sweeps up to $\mu_0 H = \pm 10$ T (Fig. 3). Although a positive magnetoresistance (PMR) shows in all the magnetotransport measurements in Fig. 1(c) and Fig. 2 between $\mu_0 H = \pm 2$ T, we observe a transition from PMR to negative magnetoresistance (NMR) between $|\mu_0 H| = 2.5$ T and 3 T. This transition field is denoted by $H^*$ in the insets of Fig. 3. A similar behavior was previously observed by Kou *et al.* on a 10 QL uniformly doped magnetic TI film [17], which is similar to the thickness of our sample. We note that this behavior is usually absent in thinner QAH samples [15–18,20,39]. This further supports our hypothesis regarding the dissipative quasihelical edge states on the side surface. Based on this observation, we speculate that the most likely reason for this slope change is a transition between different dominant dissipation mechanisms.

We next employ the theoretical model constructed by Wang *et al.* to understand the transport properties of a QAH system with both chiral and non-chiral edge states [23]. In this model, the transmission probability for a state propagating from electrode $i$ to electrode $j$ is $T_{ij}$. For a standard Hall bar with $N$ terminals, we identify $i = N +1$ as $i = 1$. Therefore, for chiral edge states in the QH effect or QAH effect we have $T_{i,i+1} = 1$ for $i = 1$ to $N$, and all other $T_{ij} = 0$. For helical edge states in the QSH effect we have $T_{i,i+1} = 1$ and $T_{i+1,i} = 1$ for $i = 1$ to $N$, and all other $T_{ij} = 0$. For quasihelical edge states, however, the transmission probability between neighboring electrodes is not perfect, hence we have $T_{i,i+1} = k_1$ and $T_{i+1,i} = k_2$ with $k_1, k_2 < 1$, and all other $T_{ij} = 0$. Consequently, the total transmission probability for a system with quasihelical edge states coexisting with chiral edge states should be given by $T_{i,i+1} = 1+ k_1$ and $T_{i+1,i} = k_2$, and all other $T_{ij} = 0$. In the presence of an external magnetic field, the backscattering of quasihelical edge states can be enhanced due to the breaking of TRS, while the chiral edge states remain robust against backscattering. When the magnetic length $l_B$ becomes smaller than the mean free path $l_{mfp}$ in an increasingly stronger magnetic field, $k_1$ and $k_2$ will start to approach zero. As a result, the influence of quasihelical edge states at high magnetic fields becomes weaker and $\rho_{xx}$ begins to decrease with increasing magnetic field after $l_B < l_{mfp}$ [23]. In Fig. 3, this transition occurs at $\mu_0 H \sim 2.5$ T and results in the kinks around $H^*$, where $l_B \sim 15.8$ nm, suggesting that $l_{mfp}$ in our system can be around 20 nm. In an even higher magnetic field beyond $H^*$, the quasihelical edge states are suppressed, and the dissipation becomes dominated by thermally activated bulk states, giving rise to the NMR in Fig. 3(a) and stabilized $\rho_{xy}$ in Fig. 3(b) due to the strong localization under high magnetic fields.

To further study this transition behavior, we take a series of magnetic field sweeps up to $\mu_0 H = \pm 5$ T at different temperatures, as shown in Fig. 4(a). To examine the $T_s$ dependence of $H^*$, the enlarged images of the downward magnetic field sweeps in Fig. 4(a) around $H^*$ are shown in Fig. 4(b). The transition initially occurs at $H^* = \mu_0 H \sim 2.5$ T at $T_s = 16$ mK. As $T_s$ increases, $H^*$ continues to shift toward lower magnetic field. The position of $H^*$ becomes harder to resolve above 400 mK because the amplitude of the kink gradually shrinks. The explanation for this phenomenon is the increasing bulk states at elevated $T_s$. Figure 4(c) shows the slope of $\rho_{xy}$-$\mu_0 H$ curves between $\mu_0 H = 0.5$ T and 1 T extracted from Fig. 4(b). In this regime, the magnetization is saturated and the $\mu_0 H$ dependence of $\rho_{xy}$ becomes linear, which can be treated as an analog of the Hall coefficient in the ordinary Hall effect [40–43]. Although it cannot be used to precisely compute the carrier concentration, it still provides qualitative information about the carrier concentration. As shown in Fig. 4(c), the absolute value of this slope increases with increasing $T_s$. In the QAH regime, this means that more surface and bulk states are thermally activated at higher $T_s$, leading to enhanced diffusive bulk transport which undermines the precision of the QAH effect and gradually dominates the influence of quasihelical edge states (see more discussions in the Supplemental Material [29]). As a result, the bulk states can outweigh the quasihelical edge states and become the dominant dissipation mechanism from a lower magnetic field at a higher temperature, resulting in the decreasing $H^*$ with increasing $T_s$ in Fig. 4(b).

Our speculation is further supported by the $\mu_0 H$ dependence of activation energy $E_a$. By extracting the $\rho_{xx}$-$T_s$ relations from Fig. 4(a), we calculate the corresponding $E_a$ at different magnetic fields using Arrhenius fits (Fig. 4(d)). We note that because of the presence of the coercive peak between $-1$ T $< \mu_0 H < 0$ T in the downward sweep (0 T $< \mu_0 H < 1$ T in the upward sweep), the data in these regimes are not used for the Arrhenius fit. The obtained $E_a$-$H$ relation is symmetric about $\mu_0 H = 0$ T, and $E_a$ is maximized at $\mu_0 H = 0$ T (see more discussions in the Supplemental Material [29]). Furthermore, in contrast to the strong magnetic field dependence at $|\mu_0 H| \leq 2$ T, $E_a$ saturates at $\sim 16$ µeV and becomes independent of $\mu_0 H$ at $|\mu_0 H| \geq 3$ T. This $E_a$-$H$ relation is consistent with our picture regarding the shift between different dominant dissipation mechanisms in the QAH effect. Specifically, because the dissipation below $H^* \sim 2.5$ T is dominated by the magnetic-field dependent quasihelical edge states, the calculated $E_a$ shows a $\mu_0 H$ dependence. Whereas beyond $H^* \sim 2.5$ T, bulk states become the major source of dissipation, resulting in nearly constant $E_a$.

In summary, we demonstrate that the residual dissipation and non-vanishing $\rho_{xx}$ observed on thick QAH samples can be ascribed to the coexistence of chiral edge states, quasihelical edge states, and bulk states. The different magnetic field dependences between bulk states and non-chiral edge states in magnetic field sweeps lead to a kink structure that corresponds to the transition in the dominant dissipation mechanism. The activation energy also exhibits a magnetic field dependence that is in good agreement with the shift of the dominant dissipation mechanism at the transition field. We expect that more quantitative analysis regarding the interaction between different dissipative states and their activation behavior are desired for achieving a QAH insulator with less residual dissipation and less demanding experimental conditions.

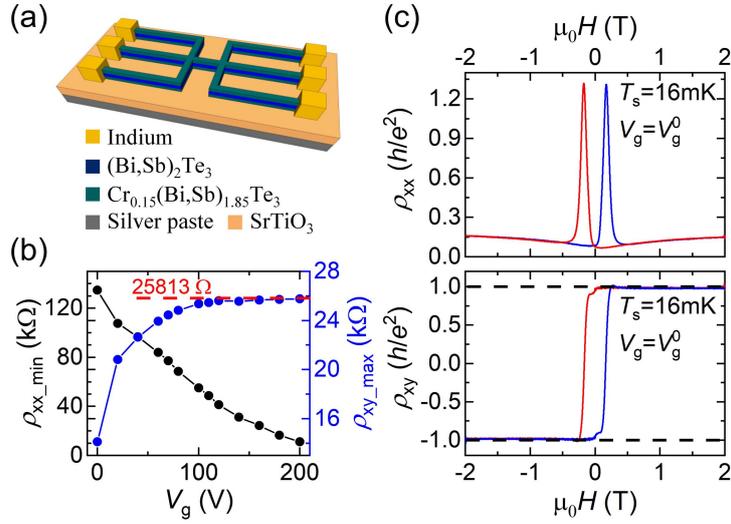

**FIG. 1** (color online). **The QAH effect in the magnetic TI sandwich heterostructure.** (a) Schematic of the Hall bar device. (b) $V_g$ dependence of the minimum $\rho_{xx}$ (black) and maximum $\rho_{xy}$ (blue) at $T_s$ = 16 mK. (c) $\mu_0H$ dependence of $\rho_{xx}$ and $\rho_{xy}$ at $T_s$ = 16 mK under $V_g = V_g^0$ = +200 V. The red and blue curves correspond to downward and upward sweeps, respectively. The dashed lines in (b) and (c) denote the quantized value $\pm h/e^2$.

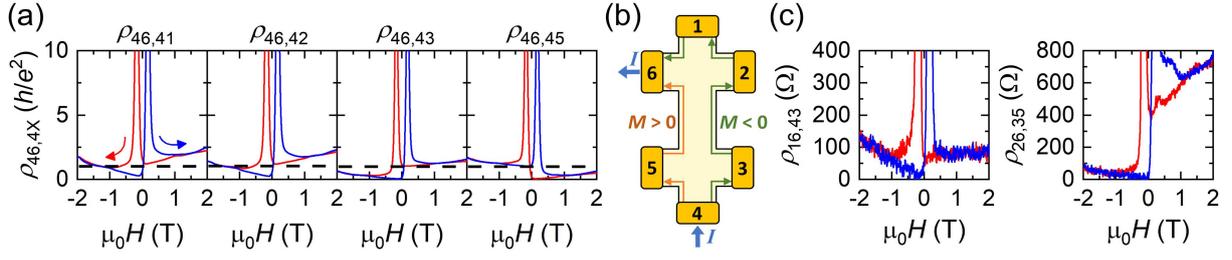

**FIG. 2** (color online). **Three-terminal and non-local measurements.** (a) $\mu_0 H$ dependence of the three-terminal resistance at $T_s = 16$ mK and $V_g = V_g^0$. $\rho_{ij,mn}$ means the resistance obtained by passing the current from electrode $i$ to electrode $j$ and measuring the voltage drop between electrode $m$ and electrode $n$. The blue and red curves represent the upward and downward sweeps, respectively. The dashed lines denote the quantized value $h/e^2$. (b) The schematic illustrating the chirality of the QAH edge states under $M > 0$ and $M < 0$ at zero external magnetic field. (c) $\mu_0 H$ dependence of the non-local resistance $\rho_{16,43}$ and $\rho_{26,35}$ taken at $T_s = 16$ mK and $V_g = V_g^0$.

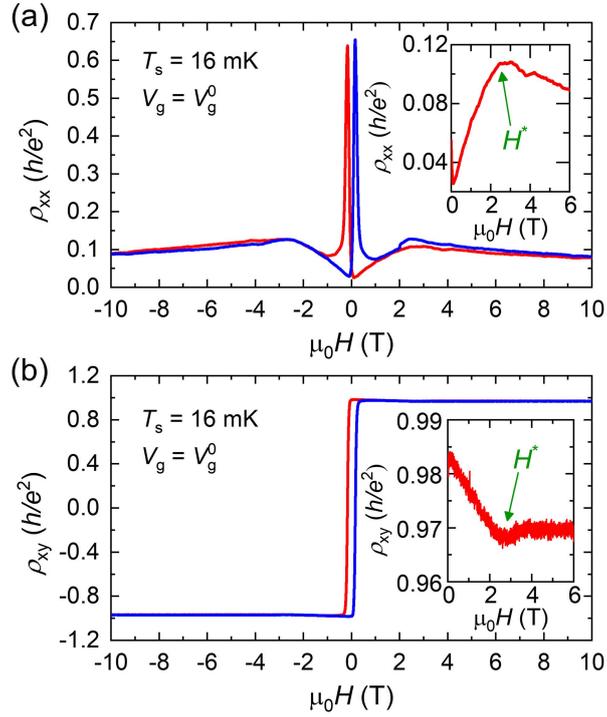

**FIG. 3** (color online). **The QAH effect under high magnetic field.** $\mu_0 H$ dependence of (a) $\rho_{xx}$ and (b) $\rho_{xy}$ up to $\mu_0 H = \pm 10$ T at $T_s = 16$ mK and $V_g = V_g^0$. The red and blue curves represent the downward and upward sweeps, respectively. The insets in (a) and (b) show the enlarged images of the downward sweeps of $\rho_{xx}$ and $\rho_{xy}$ around the point of the slope change, respectively. The transition field at which the slope change occurs is denoted by $H^*$.

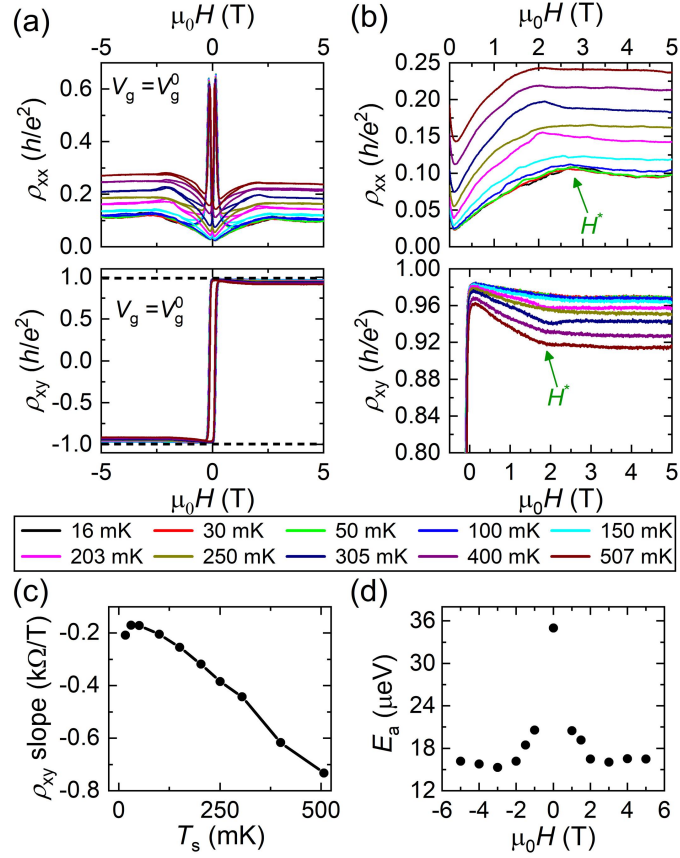

**FIG. 4** (color online). **Temperature dependence of the magnetotransport measurements.** (a) $\mu_0 H$ dependence of $\rho_{xx}$ (upper) and $\rho_{xy}$ (lower) up to $\mu_0 H = \pm 5$ T taken at different $T_s$ under $V_g = V_g^0$. A kink feature is observed both in $\rho_{xx}$ and $\rho_{xy}$ when $|\mu_0 H| \sim 2.5$ T. (b) The enlarged images of the downward magnetic field sweep in (a) around the kink structure. The position of the kink is denoted by $H^*$, which shifts toward lower magnetic field as $T_s$ increases. (c) $T_s$ dependence of the $\rho_{xy}(\mu_0 H)$ slope between $\mu_0 H = 0.5$ T and 1 T extracted from (b). (d) $\mu_0 H$ dependence of $E_a$ extracted from the $\mu_0 H$ dependence of $\rho_{xx}$ in (a).

**Acknowledgements**

This work is funded by the Engineering and Physical Sciences Research Council (EPSRC Grants No. EP/S019324/1 and No. EP/L020963/1). We thank K. X. Zhang for the helpful discussion. S.-W. W. acknowledges the support from the Taiwan-Cambridge Scholarship provided jointly by the Ministry of Education, Taiwan, and the Cambridge Trust, United Kingdom. C. Z. C. acknowledges the support from the Gordon and Betty Moore Foundation's EPiQS Initiative (Grant No. GBMF9063) and NSF-CAREER award (DMR-1847811). D. X., C. Z. C., and N. S. acknowledge the support for sample synthesis from the Penn State 2DCC-MIP under NSF Grant No. DMR-1539916.

# Supplemental Material

# Demonstration of Dissipative Quasihelical Edge Transport in Quantum Anomalous Hall Insulators

Shu-Wei Wang[1,*], Di Xiao[2], Ziwei Dou[1], Moda Cao[1], Yi-Fan Zhao[2], Nitin Samarth[2], Cui-Zu Chang[2], Malcolm R. Connolly[3], and Charles G. Smith[1]

[1]*Cavendish Laboratory, Department of Physics, University of Cambridge, Cambridge, CB3 0HE, UK*

[2]*Department of Physics, The Pennsylvania State University, University Park, Pennsylvania 16802, USA*

[3]*Department of Physics, Imperial College London, London, SW7 2AZ, UK*

*To whom correspondence should be addressed. E-mail: sww38@cam.ac.uk

This document contains:

1. Material growth and device fabrication
2. Measurement details
3. Correction for the imperfect Hall bar geometry
4. Gate dependence of the four-terminal measurements
5. Three-terminal measurements
6. Non-local measurements
7. Enhanced diffusive bulk transport at higher temperatures
8. The Arrhenius-type transport and the thermal activation energy ($E_a$)
9. The possible difference between sample temperature ($T_s$) and electron temperature ($T_e$)

# 1. Material growth and device fabrication

The growth of the Cr-doped (Bi,Sb)$_2$Te$_3$ sandwich heterostructures was carried out in a commercial EPI-620 molecular beam epitaxy (MBE) system with a base pressure lower than $2 \times 10^{-10}$ mbar. The insulating SrTiO$_3$ (111) substrates used for the growth of all the sandwich heterostructures were first soaked in 90 °C deionized water for 1.5 hours, and then annealed at 985 °C for 3 hours in a tube furnace with pure oxygen flow. Through the above steps, the surface of SrTiO$_3$ substrates were passivated and became atomically flat. These heat-treated insulating SrTiO$_3$ substrates were then outgassed at ∼ 530 °C for 1 hour before the growth of the TI heterostructures. High purity Bi (99.999%), Sb (99.9999%), Te (99.9999%), and Cr (99.999%) were evaporated from Knudsen effusion cells. The flux ratio of Te per (Bi + Sb) was set to be > 10 to minimize the Te deficiency in the TI films. Each layer of the sandwich heterostructure was grown with a different Bi/Sb ratio by adjusting the temperatures of Knudsen cells to tune the chemical potential close to its charge neutrality point. The growth rate for the films was controlled at ∼ 0.25 QL per minute, and the SrTiO$_3$ substrates were maintained at 240 °C during the sample growth. The growth process was monitored in situ by reflection high energy electron diffraction (RHEED). After the growth, the TI films were annealed at ∼ 240 °C for 30 minutes to improve the crystal quality before being cooled down to room temperature. The sharp and streaky '1×1' diffraction patterns in RHEED images, as shown in the supplementary information of Xiao *et al.* [1], indicate high crystal quality of the magnetic TI/TI/magnetic TI sandwich heterostructures.

Contamination from the environment can substantially change the properties of magnetic TIs and cause serious problems for the observation of the QAH effect [2–4]. Given the small size of the exchange gap [5–9], the environmental doping effect can easily shift the Fermi level out of the exchange gap and quench the QAH effect. Therefore, the magnetic TI device used in this study was scratched into a Hall bar by a metal tip to avoid introducing doping and contamination from the environment and lithographic processes [4,5,10,11].

The scratched Hall bar sample was then mounted onto a leadless chip carrier (LCC) using silver conductive paste. Indium pieces were manually placed onto the terminals of the Hall bars as Ohmic contacts, and these indium contacts were later connected to the bond pads of the LCC by gold wires. Finally, the inner base of the LCC, which is electrically contacted with the SrTiO$_3$ substrate by silver paste, was connected to some of the bond pads on the LCC to enable the manipulation of back-gate voltage. The stack structure of the magnetic TI sandwich sample is schematically shown in Fig. S1(a). A 5 QL undoped (Bi,Sb)$_2$Te$_3$ region is sandwiched by two 3 QL Cr-doped (Bi,Sb)$_2$Te$_3$ regions, forming a 3-5-3 TI heterostructure on the SrTiO$_3$ substrate.

Figure S1(b) shows the optical photograph of the Hall bar. The dark part is the magnetic TI film and the six silver regions are the indium contacts. The light area is the transparent SrTiO$_3$ substrate. The Hall bar was made with elongated leads to prepare for possible future scanning probe microscopy (SPM) experiments. The dimension of all the channels of the Hall bar is shown in Fig. S1(c), and the aspect ratio of the middle channel of the Hall bar is ∼ 6.5.

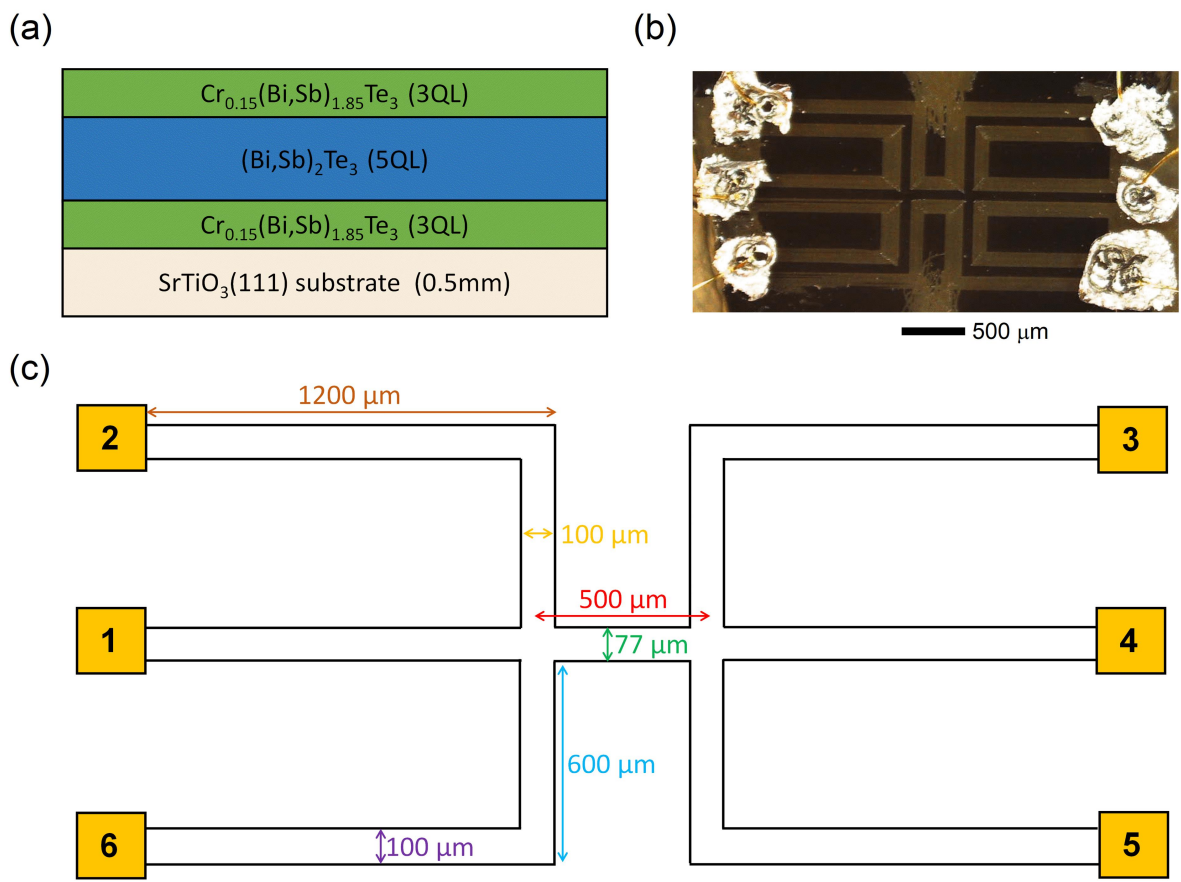

**Figure S1. The stack structure and device geometry of the sample.** (a) The layer structure of the magnetic TI sandwich sample. (b) An optical photograph of the scratched Hall bar. (c) The device layout and dimensions of the Hall bar. The aspect ratio of the middle channel is ∼ 6.5.

## 2. Measurement details

The transport measurements were performed in a Bluefors LD400 cryogen-free dilution refrigerator, which has a base temperature of 10 mK (measured at the mixing chamber stage) and a superconducting magnet that can generate up to 12 T of out-of-plane magnetic field. A separate sample thermometer was installed near the sample holder to accurately measure the actual temperature of the sample. The temperature described in the main text is the sample temperature ($T_s$) instead of the mixing chamber temperature ($T_m$).

Standard lock-in techniques and multi-terminal measurements were performed using an excitation current of 1 nA at a frequency of 15 Hz with two Zurich Instruments MFLI lock-in amplifiers and one Signal Recovery 7265 DSP lock-in amplifier. In most cases, the Stanford Research Systems SR560 voltage pre-amplifiers and the SR570 current pre-amplifier were used in the measurement circuits to reduce the noise level. The back-gate voltage for the device was supplied by a Keithley 2400 Source Measure Unit (SMU) with a maximum output of 200 V.

### 3. Correction for the imperfect Hall bar geometry

Because the Hall bar used in this study was not lithographically defined, a certain extent of misalignment between the probes measuring the longitudinal and Hall voltage difference is present, which leads to the mixing between the longitudinal resistance ($R_{xx}$) and Hall resistance ($R_{xy}$) signals. This effect is particularly serious for the $R_{xy}$ signals around the plateau-plateau transition regimes near the coercive field where the measured $R_{xy}$ will inevitably pick up the extremely large electrical signals from $R_{xx}$ in such an imperfect Hall bar geometry. Thus, a correction for the measured $R_{xy}$ is necessary. The principle behind this correction is described as follows.

Considering the ideal butterfly structure in $R_{xx}$ and square loop in $R_{xy}$, as shown in Fig. S2, the hysteresis for $R_{xx}$ and $R_{xy}$ in a magnetic field sweep should have the relations:

$$R_{xy\uparrow}(H) + R_{xy\downarrow}(-H) = 0 \tag{3.1}$$

and

$$R_{xx\uparrow}(H) = R_{xx\downarrow}(-H), \tag{3.2}$$

where the ↑ and ↓ in the subscripts denote the upward and downward magnetic field sweeps, respectively. However, for a real measurement with misaligned voltage probes, the $R_{xy}$ signal actually has a component of the $R_{xx}$ signal, and the relation becomes

$$R_{xy\uparrow}(H)_{\text{measured}} + R_{xy\downarrow}(-H)_{\text{measured}} = R_{xy\uparrow}(H) + R_{xy\downarrow}(-H) + R_{xx\uparrow}(H) + R_{xx\downarrow}(-H), \tag{3.3}$$

where the "measured" in the subscripts means the actual measured signal. Since $R_{xy\uparrow}(H) + R_{xy\downarrow}(-H)$ should equal to zero according to Eq. 3.1, the remaining part of Eq. 3.3 should be $R_{xx\uparrow}(H) + R_{xx\downarrow}(-H)$. Moreover, because $R_{xx\uparrow}(H)$ should equal to $R_{xx\downarrow}(-H)$ according to Eq. 3.2, the $R_{xx}$ components mixed in $R_{xy\uparrow}(H)_{\text{measured}}$ and $R_{xy\downarrow}(-H)_{\text{measured}}$ should be

$$[R_{xy\uparrow}(H)_{\text{measured}} + R_{xy\downarrow}(-H)_{\text{measured}}] / 2. \tag{3.4}$$

Therefore, the correction term given by Eq. 3.4 should be subtracted from both the $R_{xy}$ signals of the upward and downward magnetic field sweeps to correct the mixing of $R_{xy}$ and $R_{xx}$. It should be noted that all the $R_{xy}$ or $\rho_{xy}$ data shown in the main text and Supplemental Material are corrected, unless otherwise specified.

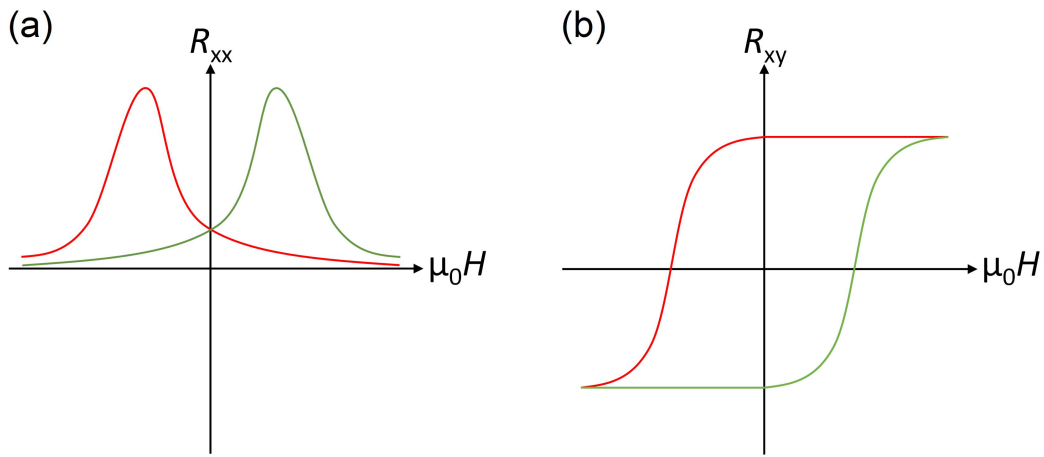

**Figure S2. Magnetic hysteresis of the longitudinal and Hall resistance.** The ideal (a) butterfly structure of $R_{xx}$ and (b) square loop of $R_{xy}$ in magnetic field sweeps. The red and green traces represent downward and upward magnetic field sweeps, respectively.

## 4. Gate dependence of the four-terminal measurements

In our previous measurements of the QAH effect on other magnetic TI devices, we noticed that the gating effect saturates at high back-gate voltage ($V_g$), and the gate sweep curves are usually hysteretic between the forward and backward $V_g$ sweeps. These two effects are possibly associated with the electrical properties of SrTiO$_3$ substrates at low temperatures. Since SrTiO$_3$ has a very high dielectric constant ($\varepsilon \sim 30000$) at low temperatures [12-14], it is commonly used as the substrate of magnetic TI devices for applying a large electric field to fine-tune the carrier density. However, after ramping $V_g$ at fridge temperatures, the SrTiO$_3$ substrate tends to be gradually charged, making the gating effect become increasingly weak. As a result, the carrier density becomes less tunable and the resistance curve in $V_g$ sweeps plateaus off at high $V_g$, which can be seen in Fig. 1(b) in the main text.

The magnetic field ($\mu_0 H$) dependence of the longitudinal resistance ($\rho_{xx}$) and Hall resistance ($\rho_{xy}$) at different $V_g$ is shown in Fig. S3, in which the $\rho_{xy}$ and $\rho_{xx}$ are obtained by measuring the $R_{14,53}$ and $R_{14,23}$ of the device, respectively, using a source-drain (S-D) current of 1 nA with an excitation frequency of 15 Hz at $T_s = 16$ mK. (Note that $R_{ij,mn}$ means the resistance obtained by passing the current from electrode $i$ to electrode $j$ and measuring the voltage drop between electrode $m$ and electrode $n$.)

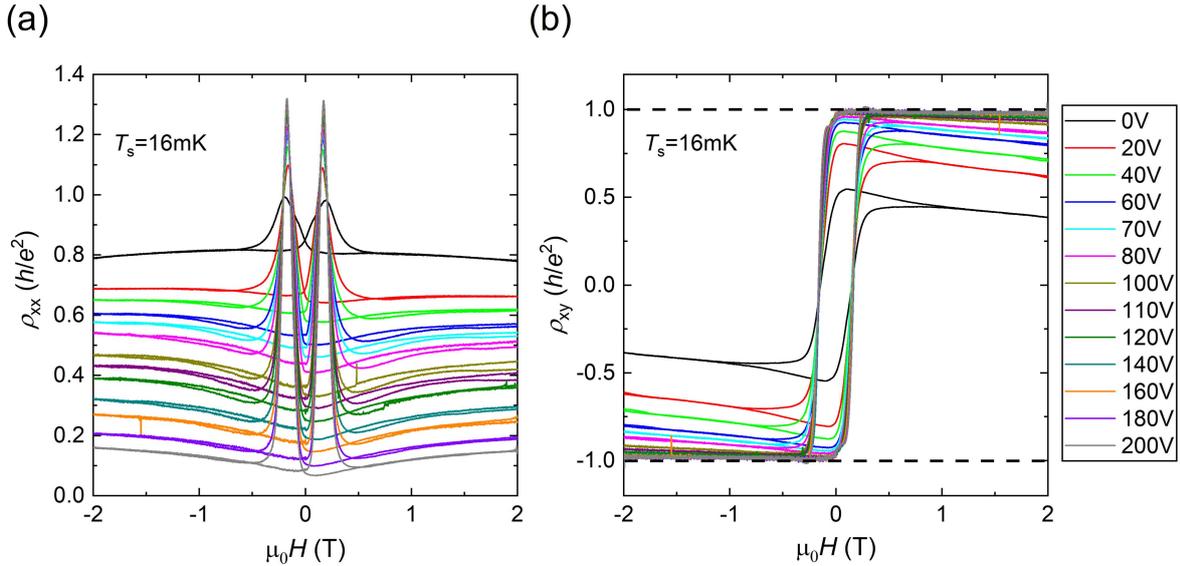

**Figure S3. The μ₀*H* dependence of ρ*xx* and ρ*xy* at different *V*g.** The magnetic field dependence of (a) $\rho_{xx}$ and (b) $\rho_{xy}$ taken at different $V_g$ using the sweep rate of 3 T/hr. The dashed lines in (b) denote the quantized value $h/e^2$.

It can be observed in Fig. S3(a) that the background level of $\rho_{xx}$ decreases with increasing $V_g$ and becomes lowest at $V_g = +200$ V, while the value of $\rho_{xx}$ at the coercive field ($H_c$) grows higher with increasing $V_g$. The different $V_g$ dependences imply that the conduction at $H_c$ is governed by a different transport mechanism to the rest of the magnetic field sweep. This is consistent with expectations because in the single domain state (i.e. when the film is fully magnetized) the transport should be dominated by edge states, whereas in the multi-domain state at $H_c$ the transport is allowed to occur through the surface. On the other hand, the two plateaux in the hysteresis loop of $\rho_{xy}$ slowly move toward the quantized value $h/e^2$ with increasing $V_g$, as shown in Fig. S3(b), suggesting a more developed QAH effect at higher $V_g$.

It is noteworthy that a small shoulder-like structure appears in the $\rho_{xy}$ curve near $\mu_0 H = 0$ T in Fig. 1(c) of the main text. Similar structures near $\mu_0 H = 0$ T have also been observed in prior studies [5,8,20]. We speculate that these small shoulder-like structures are likely a result of the heating generated in the dilution fridge and/or from the indium contacts on the samples when the orientation of the external applied magnetic field switches direction, but the exact cause is still unclear. More studies are required to clarify the origin of this heating event.

To clearly see the evolution of the $V_g$ dependence of the magnetoresistance, the minimum $\rho_{xx}$ and maximum $\rho_{xy}$ as a function of $V_g$ is plotted in Fig. S4(a), which is basically the same plot as Fig. 1(b) in the main text but with different units for the ordinate. (We note that the minimum $\rho_{xx}$ and maximum $\rho_{xy}$ are found at $\mu_0 H \sim 90$ mT before reaching zero magnetic field in the demagnetization process instead of $\mu_0 H = 0$ T. This is likely due to the demagnetization cooling effect that lowers the electron temperature and hence improves the QAH effect [5]. The minimum $\rho_{xx}$ and maximum $\rho_{xy}$ values shown in Fig. 1(b) and Fig. S4(a) are taken at these points.) It can be seen that $\rho_{xy}$ is already increasing while $\rho_{xx}$ is decreasing between $V_g = 0$ V and $+20$ V, which suggests that the device has entered the QAH regime from at least $V_g = +20$ V [4,15]. This is further supported by the evolution of $H_c$ with $V_g$ in Fig. S4(b), in which an abrupt drop in $H_c$ occurs between $V_g = 0$ V and +20 V, and then $H_c$ starts to obey an ascending trend from $V_g = +20$ V to +200 V, indicating the onset of insulator-QAH transition between 0 and +20 V. As shown in Fig. S4(a), although both the lowest $\rho_{xx} \sim 0.0667$ $h/e^2$ and the highest $\rho_{xy} \sim 0.998$ $h/e^2$ appear at $V_g = +200$ V, the trend shows that $\rho_{xx}$ is still descending and $\rho_{xy}$ is still ascending even at +200 V, suggesting that the charge neutrality point may be somewhere beyond $V_g = +200$ V. As explained in the beginning of this section, the gating effect through SrTiO$_3$ substrate tends to become weaker with the increasingly higher $V_g$ owing to the charging effect. Therefore, we did not ramp $V_g$ beyond +200 V. Moreover, since $\rho_{xy}$ has started to show a sign of saturation above $V_g = +120$ V, we speculate that $\rho_{xy}$ should not change much beyond +200 V. Nevertheless, the nearly quantized $\rho_{xy}$ and the large Hall angle $\sim 86.2º$ indicate that the Fermi level of our sample is very close to the charge neutrality point. In addition, since the QAH sample used in our study is a magnetic TI sandwich heterostructure with a thickness of $\sim 11$ nm, the helical side surface states will inevitably appear [20,21], and $\rho_{xx}$ is unlikely to vanish even if the Fermi level can be tuned to the charge neutrality point. Hence, there should be no qualitative difference for our study whether we could further tune $V_g$ to reach the minimum of the $\rho_{xx}$-$V_g$ curve in Fig. 1(b) or not.

The slopes of $\rho_{xx}$ and $\rho_{xy}$ shown in Fig. S4(c) are calculated from the data between $\mu_0 H = -1.5$ T and $-2$ T, where the magnetization of the film is saturated and the magnetoresistance in the downward sweep starts to retrace that in the upward sweep. The slope of $\rho_{xy}$ in this regime is treated as an analogue of the Hall coefficient in the ordinary Hall effect [11,15-17]. As explained in the main text, although it cannot be used to precisely compute the carrier concentration as a true Hall effect, it still provides some qualitative information about the carrier concentration. It can be seen that, after entering the QAH regime at $V_g = +20$ V, the absolute value of the $\rho_{xy}$ slope generally decreases with increasing $V_g$, implying that the 2D surface and 3D bulk states are being slowly depleted in this process. The $\rho_{xx}$ slope, on the other hand, does not have this layer of meaning, but we notice that it shows a sign change at $V_g = +20$ V, which can be the signal of the QAH phase transition and is consistent with our observations in Figs. S4(a) and (b). Finally, we observe that the longitudinal resistance at the coercive field, $\rho_{xx}(H_c)$, increases with increasing $V_g$, as shown in Fig. S4(d). The origin of this dependence can be explained by the different scattering mechanisms inside and outside the QAH regime. When the film is not in the QAH regime, 2D surface states or even some 3D residual bulk states can travel between the source and drain. The two peaks at the coercive field are caused by the scattering between these states and the magnetic domains during the propagation process [18,19]. However, when the Fermi level is tuned into the exchange gap and the film enters the QAH regime, the 2D and 3D states are mostly depleted and the two peaks in longitudinal resistance at the coercive field occur mainly via the backscattering of chiral edge states meandering through the network of domain walls, resulting in a significantly lower electrical conductivity and thus a higher resistivity at the coercive field. Hence, the $V_g$ dependence of $\rho_{xx}(H_c)$ also agrees with the previous observations about the QAH phase transition.

Therefore, although the Fermi level cannot be tuned to the charge neutrality point, a variety of evidence has verified the insulator-QAH transition between $V_g = 0$ V and +20 V, and the quantized $\rho_{xy}$ is also observed. Nevertheless, there is still a non-vanishing $\rho_{xx} \sim 0.0667\ h/e^2$, which indicates the presence of dissipative states at $V_g = +200$ V.

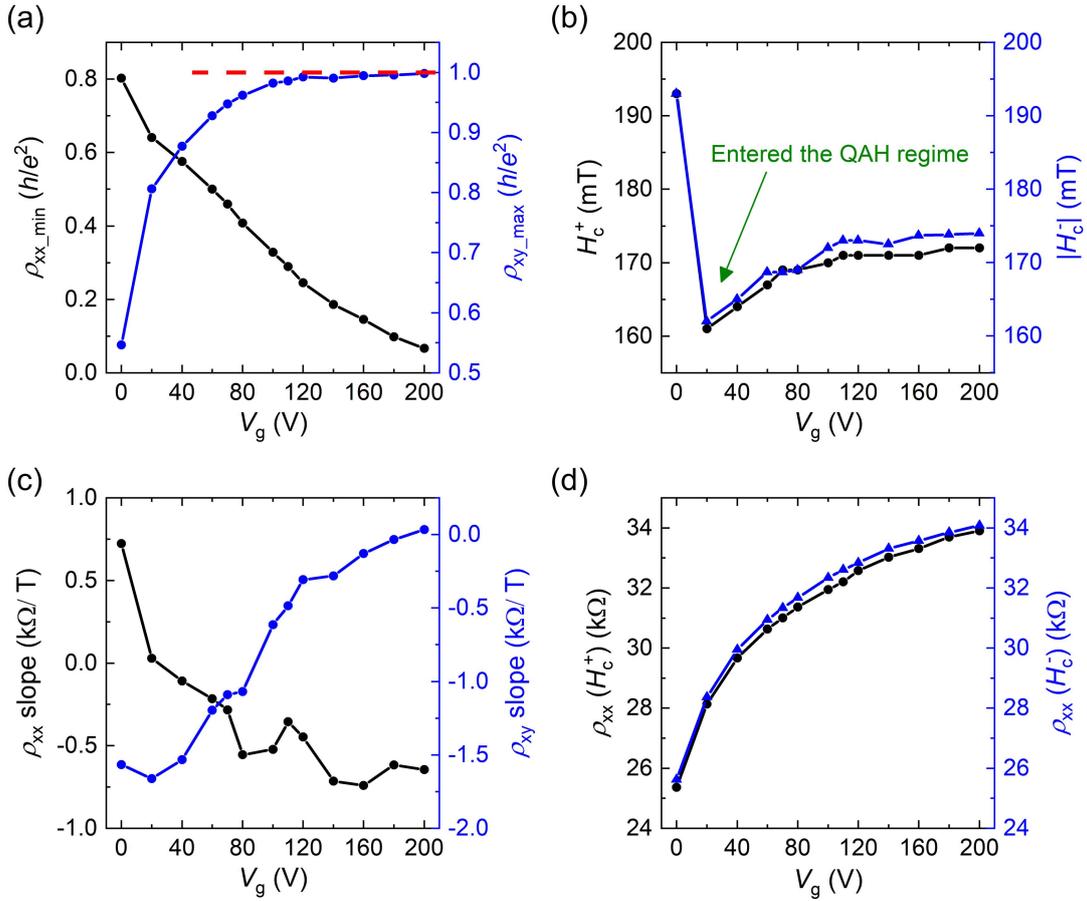

**Figure S4**. **The $V_g$ dependence of the magnetoresistance extracted from Fig. S3.** The $V_g$ dependence of the magnetoresistance extracted from Fig. S3. (a) The $V_g$ dependence of the minimum $\rho_{xx}$ and maximum $\rho_{xy}$. The red dashed line denotes the quantized value $h/e^2$. (b) The $V_g$ dependence of $H_c$. The $H_c^+$ (black) and $H_c^-$ (blue) are the coercive field values in the positive and negative magnetic field regimes, respectively. (c) The slope of $\rho_{xx}$ and $\rho_{xy}$ calculated from the data between $\mu_0 H = -1.5$ T and $-2$ T in the magnetic field sweeps. (d) The value of $\rho_{xx}$ at $H_c$ as a function of $V_g$.

## 5. Three-terminal measurements

Since the local measurements do not allow us to probe the properties of edge states, to determine the chirality of edge states, we performed three-terminal measurements, as described in Figs. 2(a) and (b). In Fig. S5, we show the full-scale plots of the three-terminal measurements for all the four different configurations, in which the slope in the magnetoresistance showing a deviation from $\rho_{xx} = 0$ and $\rho_{xy} = h/e^2$ can be clearly seen in both the positive and negative fully magnetized regimes. We note that the contact resistance of the sharing electrode 4 is not subtracted in our three-terminal measurements. As shown in Fig. S1(b), we use mm-size indium dots as the electrical contacts, and our prior studies show that the contact resistance of these indium dots is negligible [6].

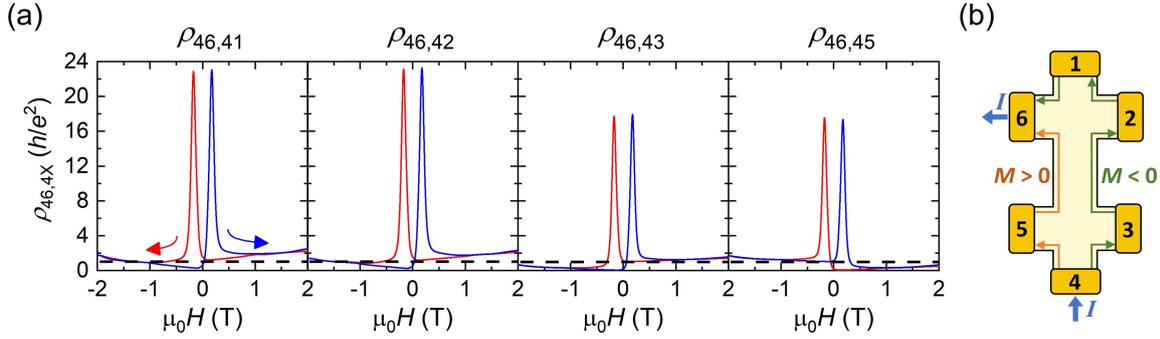

**Figure S5**. **Three-terminal measurements.** (a) The full-scale plots of the three-terminal measurements shown in Fig. 2(a) in the main text. The blue and red curves represent the upward and downward magnetic field sweeps, respectively. The dashed lines denote the quantized value $h/e^2$. (b) The schematic diagram illustrating the chirality of edge states in $M > 0$ and $M < 0$ states at zero external magnetic field.

## 6. Non-local measurements

The upper panels of Figs. S6(a) and (b) are the same plots as the magnetic field dependence of $\rho_{16,43}$ and $\rho_{26,35}$ in Fig. 2(c), while the lower panels of Figs. S6(a) and (b) show the corresponding full-scale plots of them. As described in the main text, the presence of hysteresis in the magnetic field sweeps of these non-local configurations implies the existence of dissipative edge channels [6,20]. This is reaffirmed by the observation of positive magnetoresistance (PMR) in the fully magnetized regime in these non-local measurements (Fig. 2(c)), which is different from the expected negative magnetoresistance (NMR) behavior of bulk states caused by the localization effect under strong external magnetic fields, and again suggests the existence of dissipative edge channels.

The origin of these non-chiral edge states is the topological surface states on the side surfaces of a TI sample. If the TI sample is made thin enough, the topological surface states on the side surfaces will be gapped due to the quantum confinement effect and become quasi-1D channels. Therefore, for a sufficiently thin QAH sample, the transport should be dominated by the chiral edge conduction. However, for thicker QAH samples, the quantum confinement effect on the side surfaces is not strong enough and the side surface states may not be gapped [21]. The helical nature of the topological surface states thus will become prominent in transport measurements. Furthermore, for the side surfaces the magnetization is in-plane, which can shift the position of the Dirac point in $k$-space, and the Dirac cone will no longer be protected by time-reversal symmetry [11,20]. As a result, the helical surface states on the side surfaces will no longer be immune to backscattering and will become the dissipative "quasihelical" edge states. Since the magnetic TI sandwich sample used in this work is 11 QL, the quasihelical edge states on the side surfaces are highly likely to exist and affect the theoretically dissipationless transport in the QAH effect [6,20]. The non-adiabatic transport in the QAH effect caused by these quasihelical edge states was theoretically predicted and proposed as the possible source of dissipation observed in experiments [6,20,21]. Nevertheless, a systematic study on the properties of these non-chiral edge states is still lacking. Thus, probing the behavior of the non-chiral edge states in the QAH effect is one of the major goals in our experiments.

Finally, as can be seen in Fig. S6(b), the butterfly-shaped hysteresis exhibits a double-peak structure at $H_c$, which should be attributed to the mixing between $\rho_{26,35}$ and the Hall resistance in this measurement configuration. The magnetic field dependences of $\rho_{26,35}$ and $\rho_{26,41}$ are plotted together in Fig. S6(c), where $\rho_{26,41}$ in this configuration is equivalent to the Hall resistance. The thin curves are $\rho_{26,35}$, which is the same as the lower panel of Fig. S6(b). The thick curves and dashed curves represent the corrected and raw signals of $\rho_{26,41}$, respectively. The double peak structure in $\rho_{26,35}$ overlaps with the plateau-plateau transition regime in $\rho_{26,41}$ and creates two dips in the raw signal of $\rho_{26,41}$, indicating the mixing between $\rho_{26,35}$ and $\rho_{26,41}$. In the corrected $\rho_{26,41}$ signal, the component of $\rho_{26,35}$ is removed and $\rho_{26,41}$ becomes a typical square-shaped hysteresis loop commonly seen in the Hall resistance in the QAH effect.

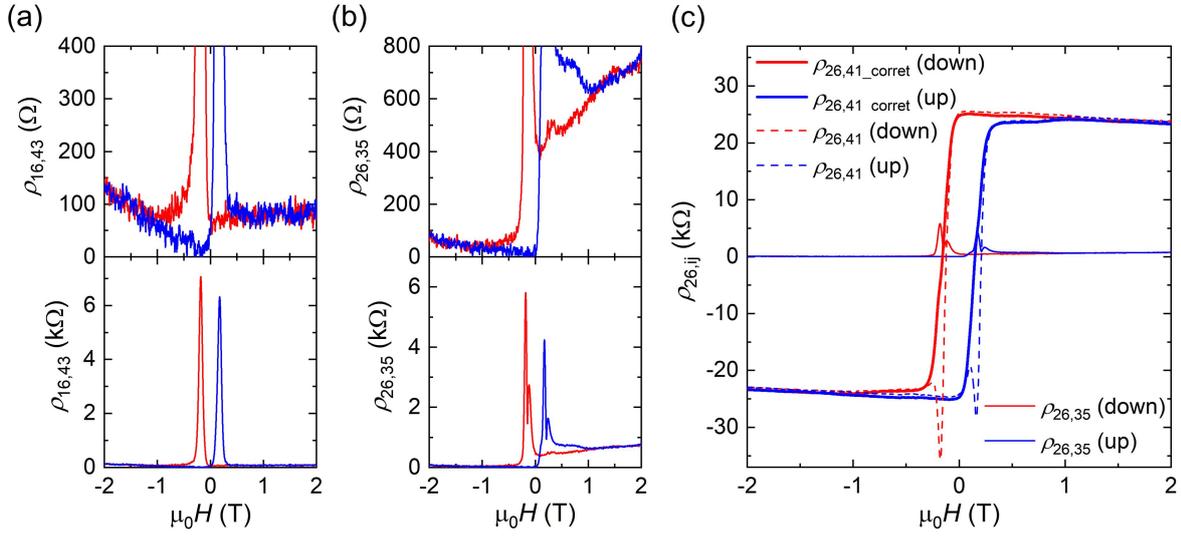

**Figure S6. Non-local measurements.** The magnetic field dependence of the non-local resistance $\rho_{16,43}$ and $\rho_{26,35}$ taken at $T_s$ = 16 mK with $V_g$ = +200 V using a sweep rate of 3 T/hr. (a) The upper panel shows the hysteresis of $\rho_{16,43}$ in the magnetic field sweeps. The lower panel is the full-scale plot of the upper panel. (b) The upper panel shows the hysteresis of $\rho_{26,35}$ in the magnetic field sweeps. The lower panel is the full-scale plot of the upper panel. A double peak structure can be observed at $H_c$, which is caused by the mixing with the Hall resistance signal. (c) The magnetic field dependence of $\rho_{26,35}$ and $\rho_{26,41}$. $\rho_{26,41}$ in this configuration is equivalent to the Hall resistance. The thin curves are $\rho_{26,35}$ as shown in the lower panel of (b). The thick curves and dashed curves represent the corrected and raw signals of $\rho_{26,41}$, respectively.

## 7. Enhanced diffusive bulk transport at higher temperatures

In Fig. 4, we discussed the $T_s$ dependence of the transition field ($H^*$). The "kink" feature related to $H^*$ actually can be seen in both $\rho_{xx}$ and $\rho_{xy}$ signals, and, though less obvious in $\rho_{xy}$, the position of $H^*$ shifts toward $\mu_0 H = 0$ T as $T_s$ increases. At first glance, it seems inconsistent with our hypothesis since $l_{mfp}$ should decrease with increasing $T_s$, and thus $H^*$ should shift away from 0 T because a stronger magnetic field is required to satisfy the condition of $l_B < l_{mfp}$. However, it is important to keep in mind that the quasihelical edge state is not the only channel affected by the temperature. As we increase $T_s$, more 2D surface and 3D bulk states as well as mid-gap states will be thermally activated, which can lead to enhanced diffusive transport and, to some extent, short the edge conduction. This is supported by the temperature dependence of the "Hall coefficient" (i.e. the slope of the $\rho_{xy}$-$\mu_0 H$ curve) as illustrated in Fig. 4(c) in the main text. Below we present more evidence to strengthen this argument.

Figure S7 shows the $T_s$ dependence of the values of $H_c$ and $\rho_{xx}(H_c)$ extracted from Fig. 4(a), in which both $H_c$ and $\rho_{xx}(H_c)$ drop with increasing $T_s$. Due to the similar reasons used for explaining Figs. S4(b) and (d), such a dependence suggests the increased carrier concentration of 2D surface and 3D bulk states at higher $T_s$. In other words, since the entire magnetic field dependence of $\rho_{xx}$ and $\rho_{xy}$ should always be considered as an outcome of the mixed dissipative edge and bulk conduction on top of the dissipationless chiral edge transport, when the weight of edge transport in conduction slowly decreases with increasing $T_s$, $\rho_{xx}(H_c)$ turns out to become smaller due to the enhanced diffusive transport.

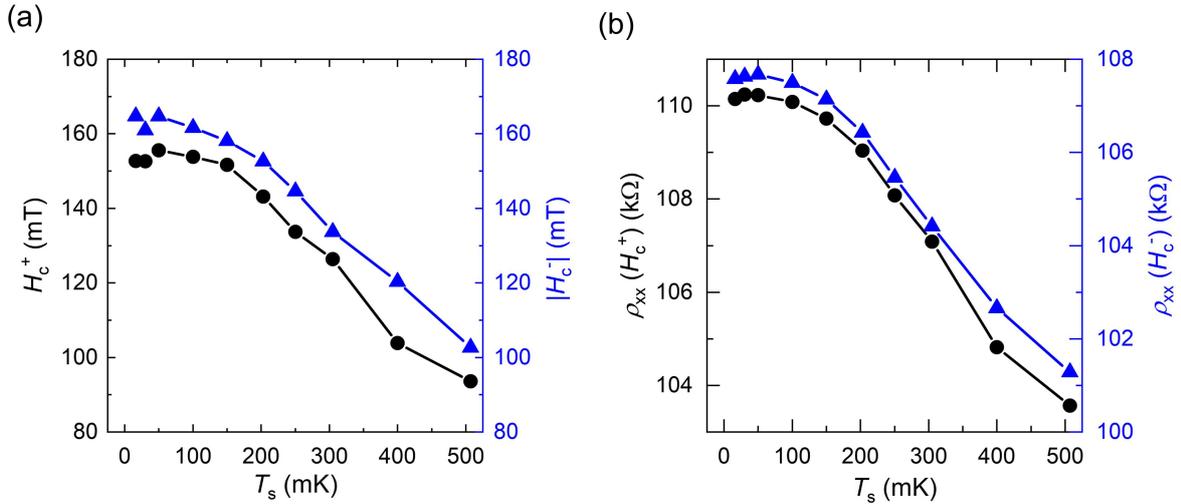

**Figure S7. The $T_s$ dependence of the coercivity and magnetoresistance.** The $T_s$ dependence of (a) $H_c$ and (b) $\rho_{xx}(H_c)$ extracted from the magnetic field sweeps of $\rho_{xx}$ in Fig. 4(a). The $H_c^+$ (black) and $H_c^-$ (blue) are the coercive field values in the positive and negative magnetic field regimes, respectively.

In conclusion, the change of carrier concentration of surface and bulk states can be probed not only by the slope of $\rho_{xy}$-$\mu_0H$ curve in the fully magnetized regime, but also can be deduced from the values of $H_c$ and $\rho_{xx}(H_c)$. Moreover, the number of surface and bulk states in the sample can be reduced by either lowering $T_s$ or tuning the Fermi level closer to the charge neutrality point by adjusting $V_g$. Therefore, by comparing the $V_g$ dependence and the $T_s$ dependence of $H_c$, $\rho_{xx}(H_c)$, and the slope of $\rho_{xy}$-$\mu_0H$ curve, we can confirm that the shift of $H^*$ toward $\mu_0H = 0$ T as $T_s$ increases is caused by the enhanced diffusive transport at elevated $T_s$. This in turn supports our hypothesis that the kinks at $H^*$ arise from the transition between edge-dominant and bulk-dominant dissipation mechanisms.

## 8. The Arrhenius-type transport and the thermal activation energy ($E_a$)

By extracting the values of $\rho_{xx}$ from the magnetic field sweeps at different $T_s$ in Fig. 4(a), the relation between $\rho_{xx}$ and $T_s$ at different magnetic fields can be obtained, as shown in Fig. S8. The longitudinal conductivity $\sigma_{xx}$ can be calculated by the tensor $\sigma_{xx} = \rho_{xx}^2 / (\rho_{xx}^2 + \rho_{xy}^2)$, and the relation between $\sigma_{xx}$ and $T_s$ at different magnetic fields is shown in Fig. S9. It can be seen that the $\sigma_{xx}$-$T_s$ curves in the high temperature regime (200 mK < $T_s$ < 507 mK) are linear and can be described by the Arrhenius equation

$$\sigma_{xx} = A\exp(-E_a/k_B T),$$

where the activation energy $E_a$ and the pre-factor $A$ can be determined from the slope and intercept of Arrhenius plots, respectively. The $E_a$-$H$ relation obtained by the Arrhenius fit is shown in Fig. 4(d). It is worth noting that the grey triangles in Fig. S8 and Fig. S9 are the data obtained in a separate measurement by varying $T_s$ at $\mu_0 H = 0$ T, which also exhibit a thermal activation transport behavior and the $E_a$ calculated from the Arrhenius fit is 34.6 μeV, consistent with the value ($E_a = 35.0$ μeV) calculated from the data at $\mu_0 H = 0$ T in Fig. 4(a). This reinforces the credibility of our analysis and strengthens our speculation regarding the change of predominant dissipation mechanisms according to the different $E_a$-$H$ dependences in different magnetic field regimes.

From the extracted $E_a$ value at $\mu_0 H = 0$ T, we deduce that the Fermi level is ~ 35 μeV away from the nearest band edge. Since the Fermi level may not be at the center of the exchange gap, we estimate that the size of the effective exchange gap should be larger than 70 μeV.

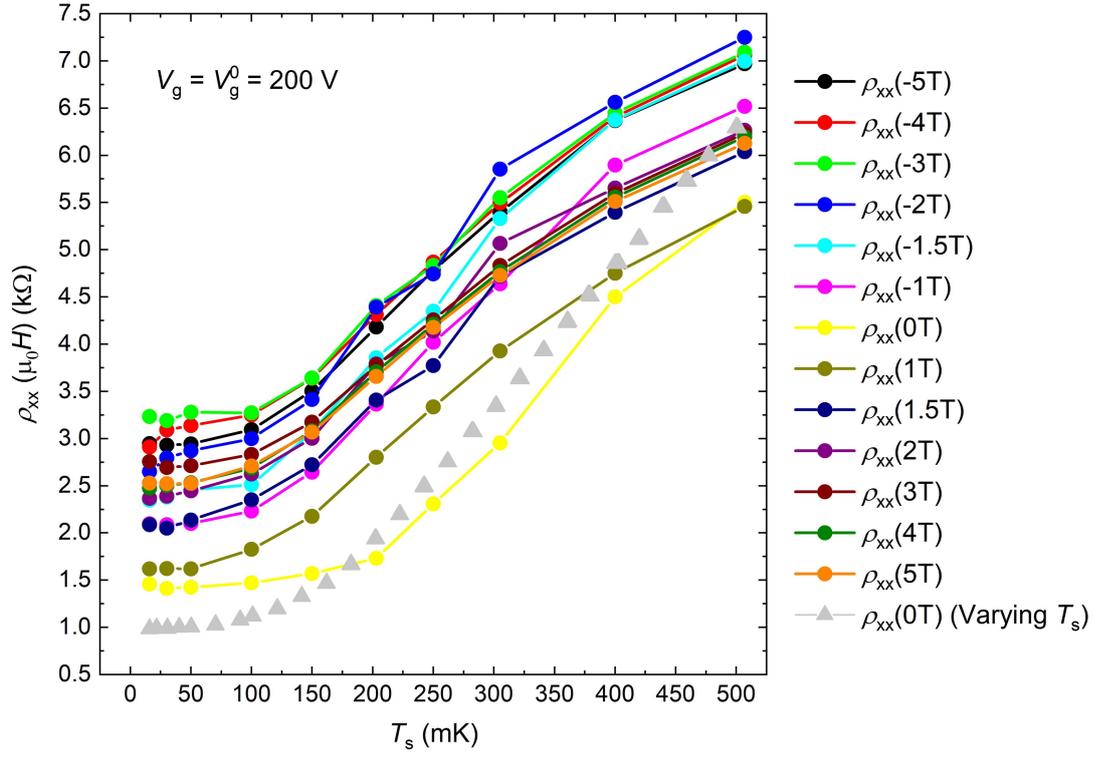

**Figure S8. The $T_s$ dependence of $\rho_{xx}$ at different magnetic fields.** The colored dots are extracted from the magnetic field sweeps of $\rho_{xx}$ in Fig. 4(a). The grey triangles are the data obtained from a separate temperature dependence measurement at $\mu_0 H = 0$ T.

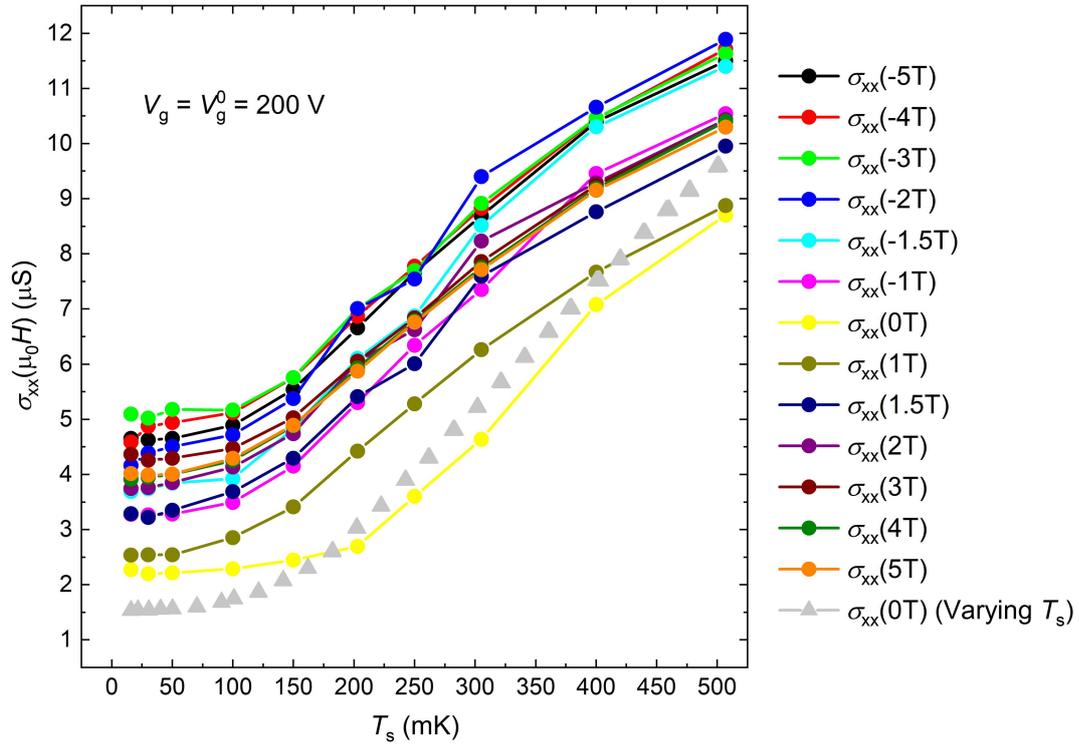

**Figure S9. The $T_s$ dependence of $\sigma_{xx}$ at different magnetic fields.** The colored dots are extracted and then converted into $\sigma_{xx}$ from the magnetic field sweeps of $\rho_{xx}$ in Fig. 4(a). The grey triangles are the data obtained from a separate temperature dependence measurement at $\mu_0 H = 0$ T.

## 9. The possible difference between sample temperature ($T_s$) and electron temperature ($T_e$)

In Figs. 4(a) and (b), we observe that the curves of $\rho_{xx}$ and $\rho_{xy}$ below $T_s = 50$ mK almost overlap and the position of $H^*$ remains at roughly the same position. Likewise, in Fig. 4(c), the slope of $\rho_{xy}$-$\mu_0H$ curve, which in some extent represents the surface and bulk carrier concentration, is very similar below $T_s = 50$ mK. Therefore, we suspect that the electrons actually might not be cooled below 50 mK in our measurements, and the sample temperature ($T_s$) read by the sample thermometer is no longer equal to the real electron temperature ($T_e$) in our experiments.

More evidence can be found in our analysis of $H_c$ and $\rho_{xx}(H_c)$ in Fig. S7 and $\rho_{xx}$-$T_s$ in Fig. S8, in which all the three physical quantities lose a clear $T_s$ dependence below $T_s = 50$ mK. Basically, both $H_c^+$ and $H_c^-$ will shift toward $\mu_0H = 0$ T when the temperature increases, as observed in our previous measurements on different magnetic TI samples [22] and in other groups' reports [4-6,10,17,20,23,24]. Nevertheless, it can be seen that $H_c$ stays at roughly the same value from $T_s = 16$ mK to 50 mK, which again implies that the electron temperature may not actually fall below 50 mK. This is further supported by the $T_s$ dependence of $\rho_{xx}$ at different magnetic fields in Fig. S7 and Fig. S8, in which the $\rho_{xx}$ values barely change when $T_s \leq 50$ mK, but become proportional to $T_s$ for $T_s > 50$ mK.

Accordingly, we speculate that $T_e$ may differ from $T_s$ when $T_s \leq 50$ mK in the experiments, and the lowest temperature that the electrons reached is possibly 50 mK.